\documentstyle[12pt]{article}
\def\GeV{\,{\rm GeV}}

\def\Gyr{\,{\rm Gyr}}

\def\Mpc{\,{\rm Mpc}}

\def\eV{{\,\rm eV}}

\def\cmm2{{\,\rm cm^{-2}}}
\def\cm2{{\,{\rm cm}^2}}
\def\cmm3{{\,{\rm cm}^{-3}}}
\def\gcmm3{{\,{\rm g\,cm^{-3}}}}
\def\kms{\,{\rm km\,s^{-1}}}

\def\mpl{{m_{\rm Pl}}}

\def\ga{\mathrel{\mathpalette\fun >}}
\def\fun#1#2{\lower3.6pt\vbox{\baselineskip0pt\lineskip.9pt
  \ialign{$\mathsurround=0pt#1\hfil##\hfil$\crcr#2\crcr\sim\crcr}}}

\begin{document}
\baselineskip=18pt
\pagestyle{empty}
\begin{center}
\bigskip

\rightline{CWRU-P6-95}
\rightline{FERMILAB--Pub--95/063-A}
\rightline{astro-ph/9504003}

\vspace{0.5in}
{\Large \bf THE COSMOLOGICAL CONSTANT IS BACK}
\vspace{0.3in}

\vspace{.2in}
Lawrence M. Krauss$^1$ and Michael S. Turner$^{2,3}$\\
\vspace{.2in}
{\it $^1$Departments of Physics and Astronomy\\
Case Western Reserve University\\
Cleveland, OH~~44106-7079}\\
\vspace{0.1in}

{\it $^2$Departments of Physics and of Astronomy \& Astrophysics\\
Enrico Fermi Institute, The University of Chicago, Chicago, IL~~60637-1433}\\

\vspace{0.1in}

{\it $^3$NASA/Fermilab Astrophysics Center\\
Fermi National Accelerator Laboratory, Batavia, IL~~60510-0500}\\
\vspace{0.2in}

({submitted to {\it Gravity Research Foundation Essay Competition}})

\end{center}

\vspace{.3in}

\centerline{\bf SUMMARY}
\bigskip
\noindent A diverse set of observations now compellingly suggest
that Universe possesses a nonzero cosmological constant.
In the context of quantum-field theory a cosmological
constant corresponds to the energy density of the
vacuum, and the wanted value for the cosmological constant
corresponds to a very tiny vacuum energy density.
We discuss future observational tests for a cosmological constant as
well as
the fundamental theoretical
challenges---and opportunities---that this poses for
particle physics and for extending our understanding
of the evolution of the Universe back to the earliest moments.

\newpage
\pagestyle{plain}
\setcounter{page}{1}
\newpage

In the early history of modern cosmology the cosmological
constant was invoked twice.  First by Einstein to obtain
static models of the Universe.\footnote{Einstein's motivation
went beyond obtaining a static solution; it was also
to insure that an empty universe satisfied Mach's principle
(see Ref.~\cite{bernstein}).}   Next by Bondi and Gold and
by Hoyle to resolve an age crisis and to construct
a Universe that satisfied the ``Perfect Cosmological Principle,''
i.e., one that appears the same at all times and places.
In both instances the motivating crisis passed and the
cosmological constant was put aside.

While Einstein called the cosmological constant his
biggest blunder and attempted to put the genie back
in the bottle, he failed.   The cosmological constant
remains a focal point of cosmology (see e.g., Refs.~\cite{tsk})
and of particle theory (see e.g., Ref.~\cite{wein}).
The former because today a wide range of observations seem to
call for a cosmological constant.  The latter because in the
context of quantum-field theory a cosmological constant corresponds
to the energy density associated with the vacuum and no known
principle demands that it vanish.

As we shall discuss, the observational case for a cosmological
constant is so compelling today that it merits consideration in spite of
its checkered history.  On the theoretical side the value of the cosmological
constant remains extremely puzzling, and it just could be
that cosmology will provide a crucial clue.
Fortunately, there are observations that should
settle the issue sooner rather than later.

What then are the data that cry out for a cosmological
constant?  They include the age of the Universe once again, the
formation of large-scale structure (galaxies, clusters of galaxies,
superclusters, voids and great walls), and the matter content
of the Universe as constrained by dynamical
estimates, Big Bang Nucleosynthesis and
X-Ray observations of clusters of galaxies.
They also relate to a bold attempt
to extend the highly successful hot big-bang model
by adding a very early epoch of rapid
expansion known as Inflation.  Inflation addresses squarely the
outstanding problems in cosmology:  the nature of the ubiquitous
dark matter and the origin of the flatness and smoothness of the
Universe as well as that of the inhomogeneity needed
to seed structure.  Inflation, which itself is based upon changes in the energy
of the vacuum,\footnote{Changes in the vacuum energy are well
understood in modern particle theory; it is the absolute scale
of vacuum energy that is poorly understood.}
predicts a spatially flat Universe and a nearly scale-invariant
spectrum of density perturbations.\footnote{Scale invariance
refers to the fact that fluctuations in the gravitational
potential are independent of scale.}  Since
big-bang nucleosynthesis precludes ordinary matter (baryons) from contributing
the mass density needed for a flat universe, inflation
requires exotic dark matter, and this has profound implications
for structure formation.  The most promising possibility
is that the bulk of the exotic dark matter is in the form
of slowly moving elementary particles left over from
the earliest moments, which leads to ``cold dark matter'' models
 for structure formation.

\vspace{0.1in}
Perhaps the most pressing piece of data mentioned above which
motivates a reconsideration of the cosmological constant involves
 the present estimate of the age of the Universe.
 The expansion age of the Universe (the extrapolated
time back to the bang) must necessarily be greater than the
age of any object within it.   Without a cosmological
constant, the expansion age is ${2\over 3}H_0^{-1}$ for a flat
(critical density) Universe and $H_0^{-1}$ for an empty Universe.
While the present expansion rate (i.e., Hubble constant $H_0$) is
still not known with precision, a variety of techniques are
converging on a value in the range $80\pm 5 \kms\Mpc^{-1}$ \cite{naturerev};
this received important support from the Hubble Space Telescope
measurement of the distance to a Virgo Cluster galaxy using
Cepheid variable stars which yielded a value of
$H_0=80\pm 17 \kms \Mpc^{-1}$ \cite{freedmanetal}.
The expansion age for a Hubble constant of $80\kms\Mpc^{-1}$ is
8.2\,Gyr for the theoretically favored flat Universe.  Even
taking a conservative lower bound to
the fraction of critical density in matter, $\Omega_{\rm
matter} \ga 0.2$, leads to an expansion age of only 10.4\,Gyr.

Therein lies the problem; the ages of the oldest globular clusters
are estimated to be $16\pm 3\Gyr$ \cite{gcages}, and
it is likely that a Gyr or so elapsed before the formation of
these stars.  The globular-cluster age estimate receives support
from other methods.  For example, studies of the cooling
of white-dwarf stars in the disk of galaxy leads to a disk age
of $9.3\pm 2\Gyr$ \cite{wd} (the disk is believed to be considerably
younger than the galaxy).

The age problem is more acute than ever before.  A cosmological
constant helps because for a given matter content and Hubble constant
the expansion age is larger.\footnote{In a universe with only matter the
expansion slows due to gravity so that $1/H_0$ is an overestimate
for the time back to the bang; a cosmological constant corresponds
to a repulsive force so that the expansion decreases more slowly
and eventually increases, leading to a larger expansion age.}
For example, for a flat universe with $\Omega_{\rm matter}= 0.2$
and $\Omega_\Lambda = 0.8$, the expansion age is $1.1H_0^{-1} = 13.2\,$Gyr
for a Hubble constant of $80\kms\Mpc^{-1}$.

Next, consider the formation of structure in the Universe.
The COBE detection of temperature variations in the cosmic
background radiation (CBR) of about $30\mu$K on the $10^\circ$ angular
scale provided striking
confirmation for the idea that structure evolved through the gravitational
amplification (Jeans' instability) of small primeval density perturbations
(variations in the density of around $10^{-5}$).
Subsequent detections of CBR anisotropy
on angular scales from about $0.5^\circ$ to $90^\circ$ by
other experiments have begun to reveal the spectrum of primeval
inhomogeneity on very-large scales\footnote{For reference, the scale
of $1\Mpc$ corresponds to galaxies, $10\Mpc$ to clusters,
$30\Mpc$ to voids, and $100\Mpc$ to the great walls}
(greater than about $100\Mpc$), and this spectrum is consistent
with that predicted by inflation \cite{wss?}.

The spectrum of density perturbations
today is not scale invariant because the Universe evolved from
an early radiation-dominated phase to a more recent matter-dominated
phase; this imposes a scale which depends upon $\Omega_{\rm matter}$,
the Hubble constant, and the amount of radiation (in the standard
scenario, the CBR and three massless neutrino species).
Through this scale, the extrapolation
from very-large scales to galaxy scales
depends upon $\Omega_{\rm matter}$ and $H_0$.
The distribution of galaxies in the Universe today can probe the
spectrum of inhomogeneity on small scales (from roughly
$1\Mpc$ to $300\Mpc$).
The agreement between the extrapolated COBE normalized
spectrum and data, including
also data on the abundance
of rich clusters, the cluster-cluster correlation function and
pairwise velocities of galaxies,
 is very good when
$\Gamma = \Omega_{\rm matter} (H_0/80\kms\Mpc^{-1})$ is
$0.3\pm 0.06$ \cite{peacock}.  This can be accomplished
with a Hubble constant of around $70-80\kms\Mpc^{-1}$ provided $\Omega_{\rm
matter}$ is around 0.3 - 0.4.  Other variants of COBE-normalized
cold dark matter that fit the data well include a very low Hubble
constant (around $30\kms\Mpc^{-1}$)
and $\Omega_{\rm matter}=1.0$, a significant increase in
the radiation level in the Universe, or the addition of a small admixture
of hot dark matter (in the form of a neutrino species of mass $5\eV$).
With the exception of the very low Hubble constant variant, which is
very much in conflict with current measurements, {\it none of these other
scenarios} are consistent with the measured age of the Universe.

Finally, consider the mass density of the Universe.
An accurate ``inventory'' of matter in the Universe is still lacking.
It is known that:  (1) most of the matter is dark,
its presence being revealed only by
its gravitational influence; (2) the fraction of critical
density contributed by ordinary matter is constrained
by big-bang nucleosynthesis to be between
$0.015(H_0/80\kms\Mpc^{-1})^{-2}$ and $0.035(H_0/80\kms\Mpc^{-1})^{-2}$
\cite{copi,kk}; (3) dynamical estimates, e.g., virial masses
of clusters of galaxies, our infall to the Virgo cluster, and
peculiar velocities of galaxies, indicate that the {\it clustered} mass
density is probably at least 20\% of critical density and
perhaps as large as the critical density.
The apparent discrepancy between the total mass
density and what ordinary matter can contribute provides the case for
exotic dark matter; the fact that few estimates indicate the clustered
mass density is as large as the critical density suggests that if the
Universe is flat, there must be an unclustered component of energy density,
like a cosmological constant.

Several authors have recently emphasized how measurements of x-rays
from rich clusters of galaxies (like Coma which contains several
thousand galaxies) together with the nucleosynthesis estimate for
$\Omega_{\rm baryon}$ can be used to estimate $\Omega_{\rm matter}$
\cite{whiteetal}.  Assuming that a rich cluster provides a
``fair sample'' of the universal mix of matter, $\Omega_{\rm matter}$
is given by the ratio of total mass to baryon mass times
$\Omega_{\rm baryon}$.  Most of the baryons in a rich cluster are in the hot,
x-ray emitting gas (as opposed to the galaxies); the x-ray
flux can be used to determine the baryonic
mass, and assuming that the gas is in virial equilibrium, the
temperature distribution of the gas determines
total mass.  Using this technique one
obtains the following estimate for the matter density:  $\Omega_{\rm matter}
(H_0/80\kms\Mpc^{-1})^{1/2} = 0.1 - 0.4$.  A
matter-dominated flat universe is only possible in this case if
the Hubble constant is extremely small, around $30\kms\Mpc^{-1}$,
or if a cosmological constant contributes the bulk of the critical
density today.

Cosmological observations thus together imply that the ``best-fit'' model
consists of matter accounting for 30\%-40\% of critical density,
a cosmological constant accounting for around 60\%-70\% of
critical density, and a Hubble constant of $70-80\kms\Mpc^{-1}$
(summarized in the Figure).  We emphasize that we are driven to this solution
by simultaneously satisfying a number of independent constraints.  Most
important in this analysis is the fact that merely violating
one of the constraints is not sufficient to allow a zero value of
the cosmological constant. {\it Unless at least two of the fundamental
observations described here are incorrect a
cosmological constant is required by the data.}

While a model with a cosmological constant
may lead to the only allowed fit to the data, and can extend our
understanding of the evolution of the Universe back the earliest
moments of the Big Bang,
 it also raises a host of fundamental concerns.
Not the least of these is the fact that a cosmological constant
implies a special epoch (today!) when for the first time
since inflation, its role in the dynamics of the Universe becomes dominant.
In the context of quantum-field theory there is the fact that
a nonzero cosmological constant corresponds to a vacuum energy
density, and particle theorists have yet to successfully constrain
its value, even to within 50 orders of magnitude of
the observational upper limit.

The energy density of the quantum vacuum receives contributions from
quantum fluctuations of arbitrarily high frequency
(and energy) and is formally infinite.
It is generally believed high-frequency fluctuations
are cutoff at the Planck scale, $\mpl \approx 10^{19}\GeV$,
and if current ideas about supersymmetry are correct, perhaps
as low as the weak scale, $1/\sqrt{G_F}
\approx 300\GeV$.  The second possibility would
lead to a vacuum energy density of around $10^{10}\GeV^4$, and
the first to a vacuum energy density of around $10^{76}\GeV^4$.  Compare these
estimates to the energy density that corresponds to the desired cosmological
constant, about $10^{-46}\GeV^4$, and to the maximum value permitted
by present data, only a factor of a few higher.  The dilemma is apparent.

The enormity of the vacuum-energy problem has led many
to conclude that there must be some kind of cancellation mechanism
at work which zeros out the ultimate value of the vacuum energy, or that
quantum-cosmological considerations favor a zero value \cite{coleman}.
However, no symmetry principle has yet been found that guarantees
a zero value for the vacuum energy, and quantum-cosmological
arguments currently rely on the shaky foundations of Euclidean
quantum gravity.  It could be then
that whatever mechanism does diminish the cosmological constant below
one's naive estimates does not
involve an exact symmetry and leaves a small vacuum energy.  It has been noted
that the desired value is close to a factor of $\exp (-2/\alpha_{\rm EM})$
less than $\mpl^4$.  (Imperfect cancellation mechanisms are not unknown;
Peccei-Quinn symmetry, which provides most
attractive solution to the strong-$CP$ problem, reduces the
electric-dipole moment of the neutron by about 20 orders of magnitude.)

Perhaps the most intriguing possibility is that the energy of
the quantum vacuum is indeed zero, but we are currently in the
midst of a phase transition where the Universe is hung up
in the false-vacuum (a mild period of inflation).  The energy scale
of this transition would correspond to $(10^{-46}\GeV^4)^{1/4} \approx
0.003\eV$, which is close to neutrino masses postulated in some
models as well as is suggested in the solution to the solar
neutrino problem.  Indeed, a model for a late-time phase transition involving
neutrino masses has been previously discussed in another context
\cite{hilletal}.

What are the possibilities for detecting a cosmological constant?
Indirectly, as we have indicated, a definitive measurement of
$H_0\ga 75\kms\Mpc^{-1}$ would necessitate a cosmological
constant or the abandonment of big-bang cosmology, and the
Hubble Space Telescope Key Project to determine $H_0$ to an accuracy of
5\% is well on its way.   More directly, one might hope to
measure the geometry of the Universe.  In particular, for a
flat Universe with a cosmological constant the distance
to an object of given redshift is much larger.\footnote{This is quantified
by the deceleration parameter which takes the value $q_0= 0.5 - 1.5
\Omega_\Lambda \sim -0.6$ rather than 0.5 for a matter-dominated flat model.}
A host of geometrical tests, including gravitational lensing \cite{gravlens},
galaxy number counts and angular size \cite{lmkdns}, offer the possibility
of detecting this difference.  It may be that the best hope lies
in CBR measurements.  In particular, for a model with a
cosmological constant the distribution of the spherical-harmonic
multipoles that characterize CBR anisotropy is distinctive,
and measurements of sufficient accuracy
are likely to be made within the next decade \cite{bunsug}.

We are currently facing a crisis in cosmology that is once again
driving us to consider the possibility that the cosmological constant is
nonzero and dominates the energy density of the Universe today.  The
challenge this poses for fundamental physics is dramatic. If, in their
third attempt to invoke a cosmological constant, cosmologists
are finally correct, the impact for our understanding of both the Universe and
of fundamental physics will be profound.

\section*{Acknowledgments}
This work was supported in part by the DOE (at Chicago,
Fermilab and Case Western Reserve) and
the NASA (at Fermilab through grant NAG 5-2788).

\end{document}